\documentclass[aps,pra,showpacs,groupedaddress,floatfix,nofootinbib]{revtex4}
\usepackage{amsmath}
\usepackage{graphicx}
\usepackage{color}

\newcommand{\beq}{\begin{eqnarray}}
\newcommand{\eeq}{\end{eqnarray}}

\begin{document}

\title{{Analytical formulas for gravitational lensing: higher order calculation}}
\author{Paolo Amore}
\email{paolo@ucol.mx}
\affiliation{Facultad de Ciencias, Universidad de Colima,\\
Bernal D\'{i}az del Castillo 340, Colima, Colima,\\
Mexico.}
\author{Santiago Arceo}
\affiliation{Facultad de Ciencias, Universidad de Colima,\\
Bernal D\'{i}az del Castillo 340, Colima, Colima,\\
Mexico.}

\author{Francisco M. Fern\'andez}\email{fernande@quimica.unlp.edu.ar}
\affiliation{INIFTA (Conicet,UNLP), Divisi\'on Qu\'imica Te\'orica, Diag. 113 y 64 S/N, \\
Sucursal 4, Casilla de Correo 16, 1900 La Plata, Argentina}

\begin{abstract}
We extend to higher order a recently published method for calculating the deflection
angle of light in a general static and spherically symmetric metric.
Since the method is convergent we obtain very accurate analytical expressions that we compare
with numerical results.
\end{abstract}

\pacs{98.62.Sb, 04.40.-b, 04.70.Bw}

\maketitle

\section{Introduction}

In a recent paper, \cite{AA06} we have introduced a new method for the calculation of the
deflection angle of light in a general static and spherically symmetric metric in
general relativity. This approach allows one to convert the integral for the angle into
a geometrically convergent series, whose terms can be analytically calculated.
Because of the strong rate of convergence of the series, typically few terms are already
sufficient to obtain highly accurate analytical results.
For this reason, previous analysis \cite{AA06} was focussed on the calculation
of first order, and a general formula, valid for an arbitrary static and  spherically
symmetric metric, was derived.  The series is not a typical perturbative expansion in
some small physical parameter, and therefore is valid also in proximity of the
``photon sphere''.

A good description of the deflection angle close to the photon sphere is particularly important for the
study of strong gravitational lensing, a topic which has received wide attention in the recent past:
for example strong gravitational  lensing in a Schwarzschid black hole has been considered by
Frittelli, Kling and Newman~\cite{FKN00} and by Virbhadra and Ellis~\cite{VE00}; Virbhadra and collaborators
have also treated the strong gravitational lensing by naked singularities~\cite{VE02} and in the presence of a
scalar field~\cite{VNC98}; Eiroa, Romero and Torres\cite{Eir02} have described Reissner-Nordstr\"om black hole lensing, while Bhadra has considered
the gravitational lensing due to the GMGHS charged black hole~\cite{Bha03}; Bozza has studied
the gravitational lensing by a spinning black hole~\cite{Boz03}; Whisker~\cite{Whi05} and
Eiroa~\cite{Eir05} have considered strong gravitational lensing by a braneworld black hole;
still Eiroa~\cite{Eir06} has recently considered the gravitational lensing by an Einstein-Born-Infeld
black hole; Sarkar and Bhadra have studied the strong gravitational lensing in the
Brans-Dicke theory\cite{SB06}; finally Perlick~\cite{Perl04} has obtained an exact gravitational lens equation
in a spherically symmetric and static spacetime and used to study lensing by a Barriola-Vilenkin monopole and
by an Ellis wormhole.

To describe the  strong gravitational lensing regime Bozza has introduced and analytical method
based on a careful description of the logarithmic divergence of the deflection angle which
allows to discriminate among different types of black holes, but which can only be used in proximity of the
photon sphere\cite{Boz02}. On the other hand, other methods, which are very precise in the weak lensing regime
have also been developed: for example, Mutka and M\"ah\"onen~\cite{Mutkaa,Mutkab} and Belorobodov~\cite{Belo02}
have derived improved formulas for the deflection angle in a Schwarzschild metric, while
Keeton and Petters\cite{Keet05} have introduced a formalism for computing corrections
to lensing observables in a static and spherically symmetric metric beyond the weak deflection limit.
These methods fail to describe the strong lensing regime and do not even reproduce a photon sphere.
This is in contrast with our method of \cite{AA06}, which already to first order predicts a photon sphere
and provides much better results than usual perturbative expansions.

The purpose of this paper is to further improve the description of the strong lensing
regime extending previous analysis \cite{AA06} to higher orders and providing with a general
formula for the systematic and analytical calculation of the deflection angle in any static
spherically symmetric metric.

The paper is organized as follows: in section~\ref{method} we review our method. In
section~\ref{ho} we obtain the general expression for the higher order contributions.
In section~\ref{res} we discuss the application of the series to a few examples.
Finally in section~\ref{concl} we draw our conclusions.

\section{The method}
\label{method}

In this section we outline the method introduced earlier \cite{AA06}.We consider
a general static and spherically symmetric metric which corresponds to the line element
\beq
ds^2 = B(r) dt^2 - A(r) dr^2 - D(r) r^2 \left(d\theta^2+ \sin^2\theta\ d\phi^2\right) \ .
\label{eq_1_1}
\eeq
Notice that the Schwarzschild metric is a special case of this equation. Although at this stage
we leave the metric coefficients $A(r)$, $B(r)$ and $D(r)$ unspecified, we assume that
$\lim_{r\rightarrow \infty} f(r)  = 1$, where $f(r) = (A(r),B(r),D(r))$ which are consistent
with a flat spacetime at infinity.

One can express the angle of deflection of light propagating in this metric in terms of the integral~\cite{Weinberg}
\beq
\Delta\phi = 2 \int_{r_0}^\infty \sqrt{A(r)/D(r)}
\sqrt{ \left[\left( \frac{r}{r_0}\right)^2 \frac{D(r)}{D(r_0)}
\frac{B(r_0)}{B(r)}-1\right]^{-1}} \frac{dr}{r} - \pi \ ,
\label{eq_1_2}
\eeq
where $r_0$ is the distance of closest approach of the light to the center of the
gravitational attraction.

In general it is possible to evaluate analytically Eq.~(\ref{eq_1_2}) only in a limited number of
cases, depending upon the form of the metric. Nonetheless we wish to prove that precise analytical
formulas can still be obtained even though no explicit expression for the exact integral may exist.

As explained in \cite{AA06} one can introduce the function
\beq
V(z) &\equiv& z^2 \frac{D(r_0/z)}{A(r_0/z)} - \frac{D^2(r_0/z) \ B(r_0)}{A(r_0/z) B(r_0/z) D(r_0)} +
\frac{B(r_0)}{D(r_0)} \ ,
\label{pot}
\eeq
where $z=r_0/r$ and thus convert Eq.~(\ref{eq_1_2}) into the form
\beq
\Delta\phi = 2 \int_0^{1} \frac{dz}{\sqrt{V(1)-V(z) }}-\pi \ .
\label{eq_1_5}
\eeq
This expression bears a close resemblance to the expression for the period of a
classical oscillator with a potential $V(z)$. It was shown in \cite{AA06} that
an efficient way of dealing with such
integral is to use the nonperturbative method based on the Linear Delta Expansion (LDE),
developed by Amore and collaborators~\cite{Am05a,Am05b}.

Let us review the main aspects of the method: we interpolate the full potential $V(z)$ with a
simpler potential $V_0(z)$, which should be chosen in such a way that the integral inside
Eq.~(\ref{eq_1_5}) can be performed explicitly when $V(z)$ is substituted by $V_0(z)$.
Then, we write
\beq
V_\delta(z) \equiv V_0(z) + \delta (V(z)-V_0(z)) \nonumber \ ,
\eeq
where $\delta$ is a dummy expansion parameter that is chosen equal to unity at the end of
the calculation. The reference potential $V_0(z)$ may depend upon a set of adjustable
parameters, which we will collectively call $\lambda$. It is sufficient for present
purposes to consider the simplest reference potential $V_0(z) = \lambda z^2$
and rewrite the deflection angle as
\beq
\Delta\phi_\delta = 2 \int_0^{1} \frac{dz}{\sqrt{E_0 -V_0(z)}} \frac{1}{\sqrt{1 + \delta \Delta(z) }}-\pi \ ,
\label{eq_1_7}
\eeq
where $E_0=V_0(1)=\lambda$,
\beq
\Delta(z) \equiv \frac{E-V(z)}{E_0-V_0(z)}-1 \ .
\label{delta}
\eeq
and $E=V(1)$.
Clearly Eq.~(\ref{eq_1_7}) reduces to the exact expression Eq.~(\ref{eq_1_5}) when $\delta = 1$.

If $|\Delta(z)|<1$ for $0\leq z \leq1$ the expansion of equation~(\ref{eq_1_7}) in powers
of $\delta$ yields a perturbation series that converges towards the exact result. It was
shown in earlier papers \cite{Am05a,Am05b,AA06} that this condition is met when $\lambda$
is greater than a critical value, $\lambda>\lambda_C$. Thus, the sequence of partial sums
converges towards the exact result that is independent $\lambda$, although each partial sum
depends on the adjustable parameter. In order to minimize the dependence of the approximate
results on $\lambda$ we resort to the the Principle of Minimal
Sensitivity (PMS)\cite{Ste81} which in present case reads:
\beq
\frac{\partial}{\partial \lambda} \Delta\phi^{(N)} = 0 \ ,
\label{eq_1_8}
\eeq
where $\Delta\phi^{(N)}$ is the partial sum of order $N$.

This condition determines an optimal value of $\lambda$ around which the truncated series
is less sensitive to changes in that parameter. In addition to it, one expects that the
resulting series exhibits improved convergence properties~\cite{AA06}. More precisely, it
has been verified that the series obtained by means of the PMS exhibit geometric rate of
convergence providing markedly accurate results with just a few terms~\cite{AA06}. In
fact, just a first order calculation proved sufficient for the prediction of the location of
the photon sphere with remarkable accuracy \cite{AA06}.

Previous calculations~\cite{AA06} also show that the rate of convergence of the PMS series
depends critically upon the physical parameters of the model, and in particular
the accuracy of the results deteriorates as $r_0$ approaches the photon sphere.
For this reason in this paper we are interested in the systematic improvement of our
results by means of general and simple formulas that facilitate
the calculation of any desired order of approximation for a given
arbitrary static and spherically symmetric metric.

\section{Higher order calculation}
\label{ho}

Under the assumption that the general metric considered here is flat at the infinity
we can express the potential in Eq.~(\ref{pot}) as
\beq
V(z) = \sum_{n=1}^\infty v_n z^n \ .
\eeq

Substitution of this expression into Eq.~(\ref{delta}) yields
\beq
\Delta(z) =  \sum_{n=1}^\infty \frac{v_n}{\lambda} \ \sum_{k=0}^{n-1}  \frac{z^k}{1+z} -1 \ .
\label{eq_1_10}
\eeq

Additionally, the angle of deflection becomes
\beq
\Delta \phi = 2 \int_0^1 \frac{dz}{\sqrt{E_0-V_0(z)}} \ \sum_{n=0}^\infty \frac{(2 n-1)!!}{2^n n!} \ \delta^n (-1)^n  \Delta^n(z) - \pi
\equiv \sum_{n=0}^\infty \Delta_n \ ,
\eeq
where the meaning of the terms $\Delta_n$ is obvious.

Assuming that $|\Delta(z)|<1$ for $z \in (0,1)$ and the series is convergent, then summation and
integration can be interchanged, thus leading to
\beq
\Delta \phi = 2 \sum_{n=0}^\infty  \sum_{j=0}^n \frac{(2 n-1)!! (-1)^{j}}{2^n j! (n-j)! \lambda^{j+1/2}} \ \delta^n
\sum_{r_1 \ r_2 \ \dots \ r_j}^\infty v_{r_1} v_{r_2} \dots v_{r_j} \ \Omega_{r_1 r_2 \dots r_j}^{(j)} - \pi
\label{defl}
\eeq
where the rank--$N$ tensors
\beq
\Omega_{n_1 n_2 \dots n_N}^{(N)} \equiv \sum_{k_1=0}^{n_1-1} \sum_{k_2=0}^{n_2-1} \dots \sum_{k_N=0}^{n_N-1} \int_0^1
\frac{z^{k_1+k_2+\dots+k_N}}{(1+z)^N} \frac{dz}{\sqrt{1-z^2}} \ .
\eeq
are dimensionless and completely symmetric in the lower indices. The explicit form of these
tensors up to order four is given in the Appendix.

By induction one derives the general expression
\beq
\Omega_{n_1 \dots n_k}^{(k)} &=& \frac{2^{k-1}}{(2k-1)!!} \sqrt{\pi} \ \left\{ \frac{\Gamma(\sum_{i=1}^k n_i/2 + 1/2)}{\Gamma(\sum_{i=1}^k n_i/2 -(k-1))}
- \sum_{P_j} \frac{\Gamma(\sum_{i\neq j}^k n_i/2 + 1/2)}{\Gamma(\sum_{i\neq j}^k n_i/2 -(k-1))} \right. \nonumber \\
&+& \left. \sum_{P_{j_1j_2}} \frac{\Gamma(\sum_{i\neq j_1 j_2}^k n_i/2 + 1/2)}{\Gamma(\sum_{i\neq j_1 j_2}^k n_i/2 -(k-1))}  + \dots
\right\} \ .
\label{omega}
\eeq
where $\sum_{P_{j_1 j_2 \dots j_l}}$ means sum of all the different combinations of the
$k-l$ elements chosen out of $k$ total elements. The amount of terms is simply given by
the binomial coefficient
$\left( \begin{array}{c} k \\ k-l \end{array}\right) = \frac{k!}{l! (k-l)!}$.

A crucial observation is that the indices of $\Omega$ take all possible values when the
sum is performed. Consequently we cannot distinguish among the possible different
combinations inside (\ref{omega}) and the same result must be obtained taking a single one
with the proper multiplicity factor given by the binomial.
We can therefore make the substitution
$\Omega_{n_1 \dots n_k}^{(k)} \rightarrow \sum_{l=1}^k \tilde{\Omega}^{(k|l)}_{n_1 \dots n_l}$
where
\beq
\tilde{\Omega}^{(k|l)}_{n1 \dots n_l} \equiv \frac{2^{k-1}}{(2k-1)!!} \sqrt{\pi} \left( \begin{array}{c} k \\ k-l \end{array}\right)
\ (-1)^{k-l} \ \frac{\Gamma(\sum_{i}^{l} n_i/2 + 1/2)}{\Gamma(\sum_{i}^{l} n_i/2 -(k-1))} \ .
\label{omegatilde}
\eeq

By substitution of this expression into Eq.~(\ref{defl}) we obtain the final expression
\beq
\Delta \phi = 2 \sum_{n=0}^\infty  \sum_{j=0}^n \frac{(2 n-1)!! (-1)^{j}}{2^n j! (n-j)! \lambda^{j+1/2}} \ \delta^n
\sum_{r_1 \ r_2 \ \dots \ r_j}^\infty v_{r_1} v_{r_2} \dots v_{r_j} \
\sum_{l=1}^j \tilde{\Omega}^{(j|l)}_{r_1 \dots r_l} - \pi \ ,
\label{general}
\eeq

In the end, we can explicitly write the expression corresponding to the first few orders as
\beq
\Delta \phi = - \pi +
\frac{315}{64 \sqrt{\lambda_{PMS} }} \Sigma_0 -\frac{105}{16 \lambda_{PMS} ^{3/2}} \Sigma_1 +\frac{189}{32 \lambda_{PMS} ^{5/2}} \Sigma_2
-\frac{45}{16 \lambda_{PMS} ^{7/2}} \Sigma_3 +\frac{35}{64 \lambda_{PMS} ^{9/2}} \Sigma_4 + \dots
\label{explicit}
\eeq
where
\beq
\Sigma_j \equiv \sum_{r_1 \ r_2 \ \dots \ r_j}^\infty v_{r_1} v_{r_2} \dots v_{r_j} \
\sum_{l=1}^j \tilde{\Omega}^{(j|l)}_{r_1 \dots r_l}  \ .
\label{sigma}
\eeq

For simplicity, in Eq.~(\ref{explicit}) we have selected the optimal parameter
of {\sl first order} $\lambda = \lambda_{PMS}$ for all orders of approximation.
The reason is that it is
not possible to obtain analytical expressions of $\lambda_{PMS}$ for greater orders, and that it
is sufficient that $\lambda_{PMS}>\lambda_C$ for the series to be geometrically
convergent. Thus, in all the applications below we choose the value of $\lambda_{PMS}$
obtained previously \cite{AA06}:
\beq
\lambda_{PMS}^{(1)} = \frac{\Sigma_1}{\Sigma_0} = \frac{2}{\sqrt{\pi}} \sum_n   v_n \frac{\Gamma(n/2+1/2)}{\Gamma(n/2)} \ .
\label{eq_1_16}
\eeq

\section{Applications}
\label{res}

In this section we consider some applications of the general formula (\ref{general}).
The first example is the Schwarzschild metric given by
\beq
B(r) = A^{-1}(r) = \left(1-\frac{2GM}{r}\right)  \ \ , \ \ D(r) = 1 \ ,
\label{eq_2_1}
\eeq
where $M$ is the Schwarzschild mass and $G$ the gravitational constant.
The angle of deflection of a ray of light reaching a minimal distance
$r_0$ from the black hole is given Eq.~(\ref{eq_1_5}), and the exact result
can be expressed in terms of incomplete elliptic integrals of the first kind\cite{Darw59}
as
\beq
\Delta\phi = 4 \sqrt{\frac{\overline{r}_0}{\Upsilon}} \ \left[ F\left(\frac{\pi}{2}, \kappa\right) -
F\left(\varphi, \kappa\right)  \right] \ ,
\label{eq_2_3}
\eeq
where $\overline{r}_0\equiv r0/GM$ and
\beq
\Upsilon \equiv \sqrt{\frac{\overline{r}_0-2}{\overline{r}_0+6}} \ \ &,& \ \
\kappa \equiv \sqrt{(\Upsilon-\overline{r}_0+6)/2\Upsilon} \ \ , \ \
\varphi \equiv \sqrt{\arcsin \left[\frac{2+\Upsilon-\overline{r}_0}{6+\Upsilon-\overline{r}_0}\right]} \ .
\label{eq_2_4}
\eeq

The ``potential'' for this metric is particularly simple since it is just a polynomial of
third degree:
\beq
V(z) = z^2 - \frac{2 GM}{r_0} z^3 \ .
\eeq

\begin{figure}
\begin{center}
\includegraphics[width=9cm]{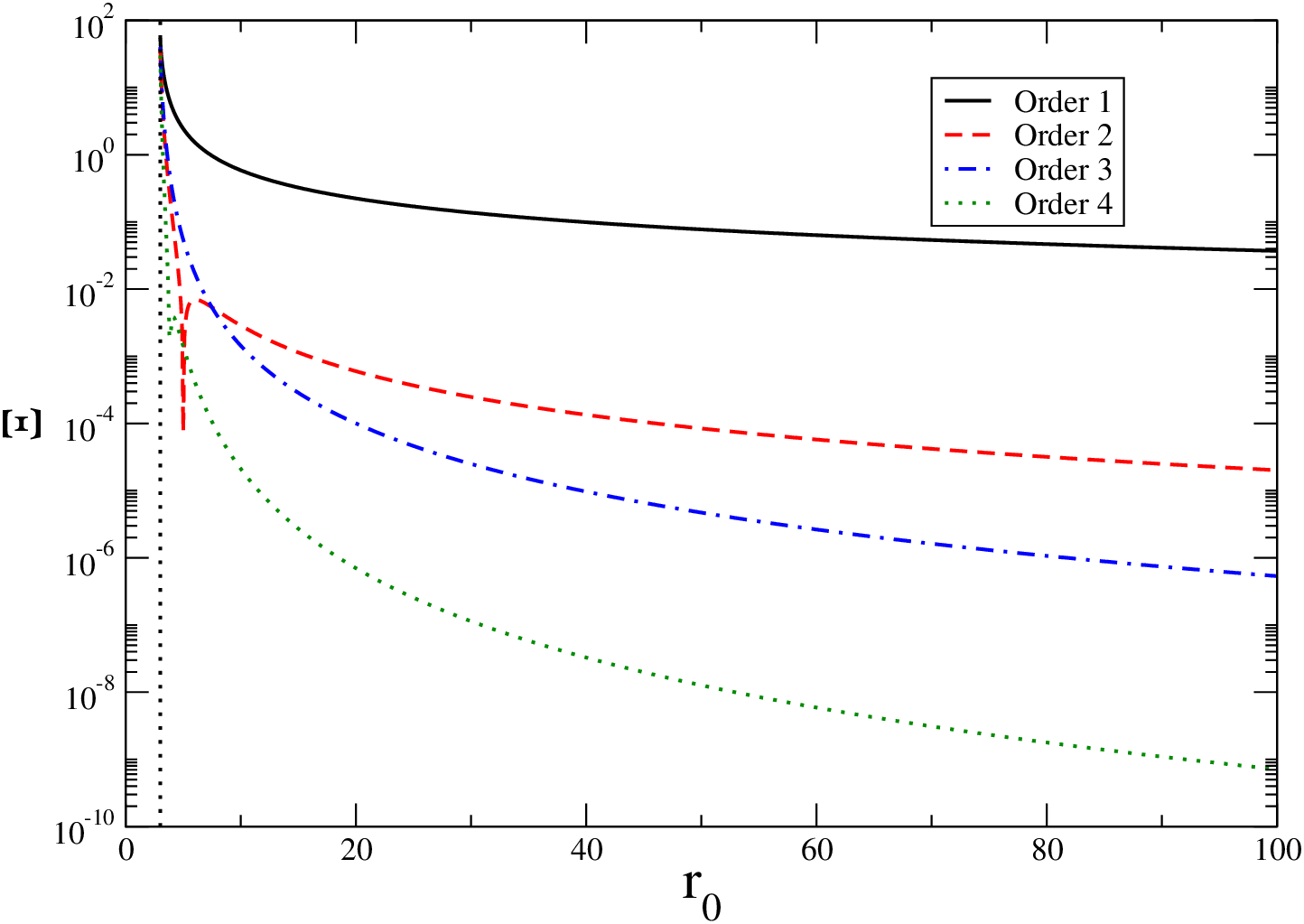}
\bigskip
\caption{Error of the deflection angle as a function of $r_0$ for the Schwartzchild metric
with $GM=1$ . The dotted vertical line shows the location of the photon sphere.
(color online)}
\bigskip
\label{FIG1}
\end{center}
\end{figure}

In order to apply Eq.~(\ref{general}) we choose $GM=1$ and calculate the error
of the deflection angle as a function of the distance of maximal approach $r_0$
($=\overline{r}_0$), defined as
\beq
\Xi = \left|\frac{\Delta\phi_{PMS} - \Delta\phi_{exact}}{\Delta\phi_{exact}} \right|
\times 100 \ ,
\eeq
where $\Delta \phi_{exact}$ is the Darwin solution~(\ref{eq_2_3}). Fig.~\ref{FIG1}
shows that the errors of the PMS approximations through fourth order are extremely
small and smaller that one percent even close to the photon sphere located at $r_0 = 3$
(the one percent error is obtained at $r_0 \approx 3.2$ for the fourth--order partial sum).

Present approach yields similar results for the Reissner-Nordstr\"om (RN) metric
\beq
B(r) = A^{-1}(r) =  \left(1-\frac{2 GM}{r}+\frac{Q^2}{r^2}\right) \  \  &,& \ \ D(r) = 1 \ .
\eeq
that describes a black hole with charge. In this case
Eiroa, Romero and Torres \cite{Eir02} have been able to express the deflection angle in
terms of elliptic integrals of the first kind (see eqn. (A3) of  \cite{Eir02}).

The ``potential'' for this example is also polynomial:
\beq
V(z) = z^2 - \frac{2 GM}{r_0} z^3 + \frac{Q^2}{r_0^2} z^4 \ .
\eeq

\begin{figure}
\begin{center}
\includegraphics[width=9cm]{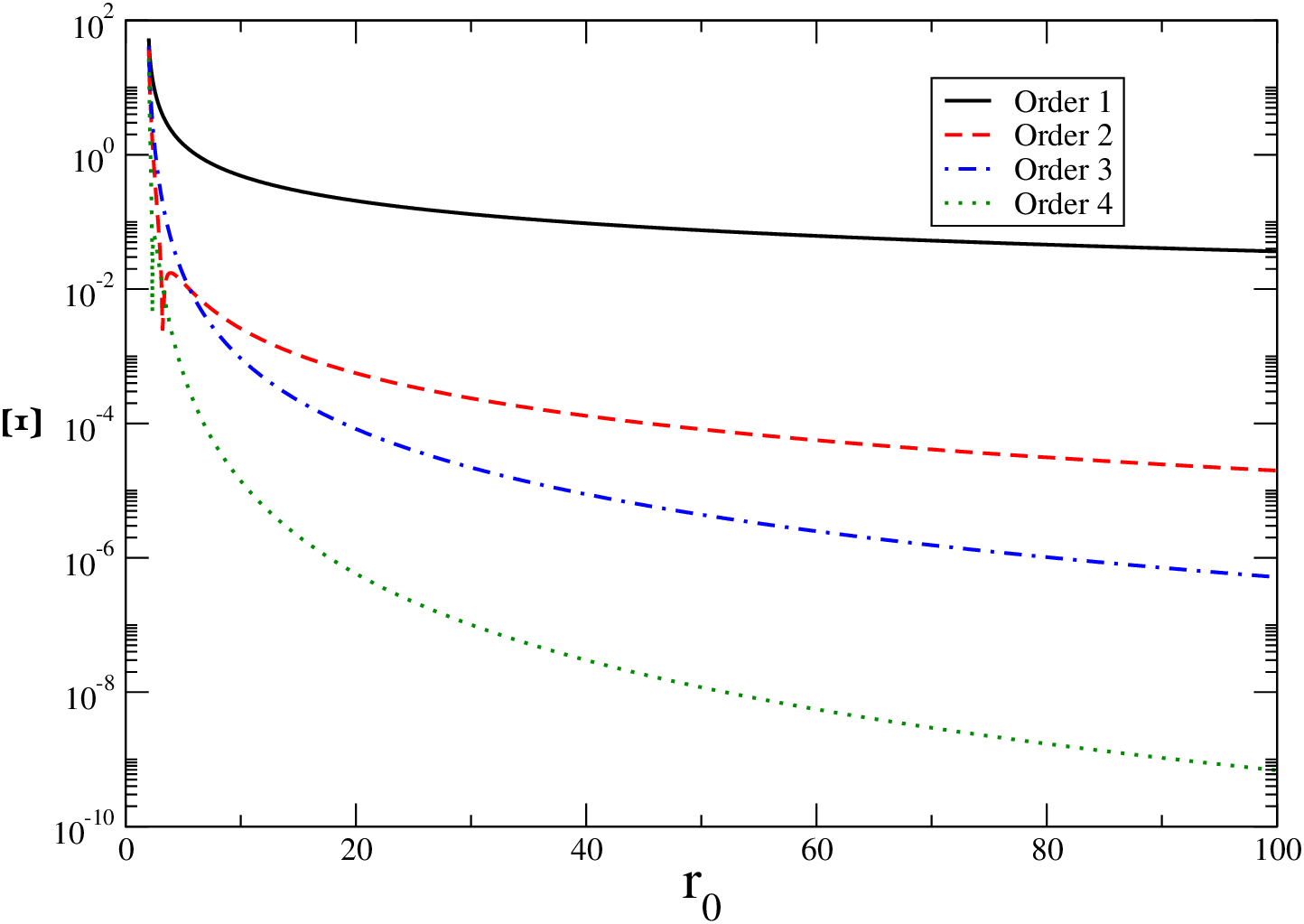}
\bigskip
\caption{Same as in Fig.~\ref{FIG1} for the RN metric with $GM=1$ and $Q=1$. (color online)}
\bigskip
\label{FIG2}
\end{center}
\end{figure}

The behavior of the error displayed in Fig.~\ref{FIG2} is similar to that observed
in the case of the Schwarzschild metric.

Both examples discussed above lead to polynomial potentials in $z$ and are amenable to
exact solutions in terms of elliptic functions. However, it should be noticed that our method
applies to more general cases that cannot be solved exactly.

In an earlier paper \cite{AA06} we considered the propagation of light in a charged black
hole coupled to Born-Infeld electrodynamics~\cite{Eir06}, which corresponds to the
effective metric
\beq
A(r) = \frac{\sqrt{\omega(r)}}{\psi(r)} \ \ , \ \ B(r) = \sqrt{\omega(r)} \psi(r) \ \ , \ \  D(r) =
\frac{1}{\sqrt{\omega(r)}} \ ,
\eeq
where
\beq
\omega(r) &=& 1 + \frac{Q^2 b^2}{r^4} \\
\psi(r)   &=& 1- 2 \frac{M}{r} + \frac{2}{3 b^2} \ \left\{ r^2 - \sqrt{r^4+b^2 Q^2} + \frac{\sqrt{|bQ|^3}}{r} \
F\left[\arccos \left( \frac{r^2-|bQ|}{r^2+|bQ|} \right) , \frac{1}{\sqrt{2}} \right] \right\}  \ .
\eeq
Here, $F(a,b)$ is the incomplete elliptic integral of first kind. We follow the convention
of Ref.~\cite{Eir06} and set $G=1$.

In this case one obtains the potential
\beq
V(z) &\equiv& z^2  \frac{\psi(r_0/z)}{ \omega(r_0/z)} - \frac{\psi(r_0) \omega(r_0)}{\omega^2(r_0/z)} +
\psi(r_0) \omega(r_0)
\eeq
which can be expanded around $z=0$ as
\beq
V(z) = \sum_{n=2}^\infty v_n z^n \ .
\label{eq_series}
\eeq

\begin{figure}
\begin{center}
\includegraphics[width=9cm]{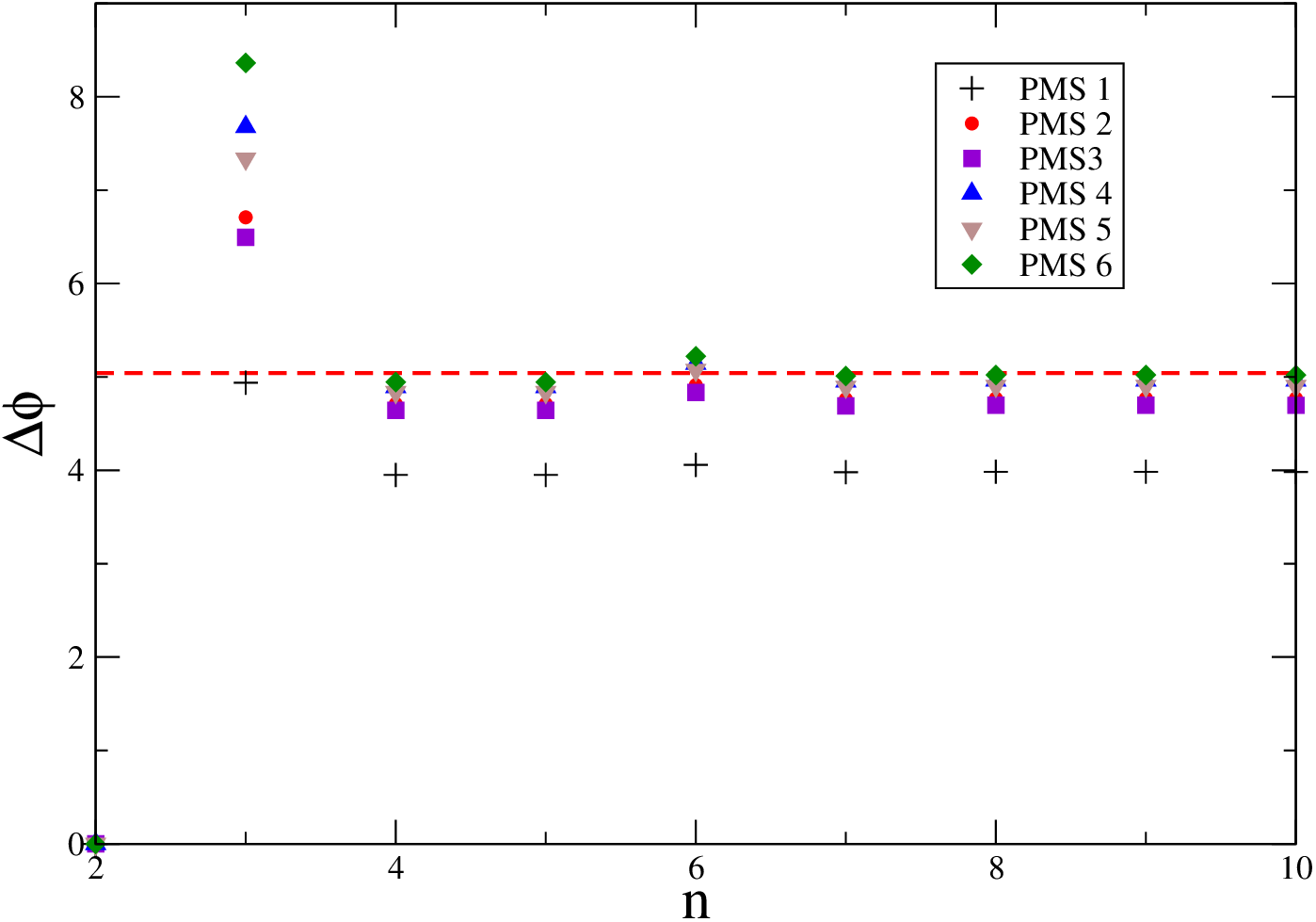}
\bigskip
\caption{Deflection angle for the Born Infeld metric with $GM=1$, $Q=1/2$, $b=1$,
and $r_0=3$. The horizontal axis specifies the order of the Taylor approximation to
the potential. The dashed line is the numerical result.
(color online)}
\bigskip
\label{FIG3}
\end{center}
\end{figure}

Clearly, in this case the accuracy of our results will depend not only upon the order
of the partial sum but also on the number of Taylor coefficients used to
approximate the potential. We illustrate this point in Fig.~\ref{FIG3} for the
deflection angle when  $GM=1$, $Q=1/2$ and $b=1$, and for $r_0 = 3$,
which is very close to the photon sphere where the angle diverges.
This situation clearly corresponds to a strong gravitational lensing regime.
The horizontal axis shows the number of Taylor coefficients of the potential used
in the calculation; the reader will appreciate that just a small number of them
($n \approx 4$) is sufficient for reasonable accuracy. For larger values of $n$
the results do not appear to improve noticeably and a sort of plateau is reached.
We point out that the main advantage of our approach is that it leads to relatively
simple analytical functions of the parameters of the metric.

Table~\ref{tab2} shows the deflection angles in terms of $r_0$ for the Born-Infeld metric
with the same parameters as before. The potential $V(z)$ is approximated
with a Taylor polynomial of order $10$ and the calculations are performed with our method
through order $4$.

\begin{table}
\caption{\label{tab2} Deflection angle for the Born-Infeld metric with $GM=1$, $Q=1/2$
and $b=1$.}
\begin{ruledtabular}
\begin{tabular}{cccccc}
$r_0$ & Numerical & $PMS_1$  & $PMS_2$ & $PMS_3$  & $PMS_4$ \\
\hline
3  &    5.04066558 &   3.9836332  &   4.769690116 &    4.695977905  &   4.961794995 \\
4  &   1.989492408   &  1.905459664  &   1.988326336  &   1.985415327  &   1.989578734 \\
5   &  1.302767649   &  1.274857482   &  1.302841883    & 1.302239811   &  1.302781935 \\
10   &  0.4908767995   &  0.4881229478  &   0.4908905773   &  0.4908704512  &   0.4908768924\\
50   &  0.08300041129   &  0.08293664219   &  0.08300048071   &  0.08300040744  &   0.0830004113\\
100  &   0.04073430348  &   0.04071929339   &  0.04073431165  &   0.04073430326     &  0.04073430348\\
\end{tabular}
\end{ruledtabular}
\end{table}

As a final example we consider the metric of Weyl gravity~\cite{MK89,Ed98,Pir04a,Pir04b}:
\beq
B(r) = A^{-1}(r) = \left(1-\frac{2\beta}{r} + \gamma r - k r^2\right)  \ \ , \ \ D(r) = 1 \ ,
\eeq
where $\beta$,$\gamma$ and $k$ are constants. When $\beta=GM$ and $\gamma$ and $k$ small enough
one recovers the Schwarzschild metric on a certain distance scale. The linear and quadratic terms
are significant only on galactic and cosmological scales, if the corresponding parameters are sufficiently
small.

The potential corresponding to this metric is once again polynomial in $z$:
\beq
V(z) = -k r_0^2 +\gamma r_0 z +z^2 -\frac{2 \beta}{r_0} z^3 \ .
\eeq
Notice that the constant term cancels out of the integral for the deflection
angle that will therefore be independent of $k$.

On defining $\Lambda \equiv \sqrt{1 - \frac{8 \beta}{\pi r_0} + \frac{2\gamma r_0}{\pi} }$
we obtain the fourth--order expression
\beq
\Delta\phi^{(4)} = - \pi + \sum_{n=0}^4 \frac{d_{2 n+1}}{\Lambda^{2 n+1}} \ ,
\eeq
where
\beq
d_1 &=& 3.19285 \\
d_3 &=& \frac{0.42847 \beta}{r_0}- 0.10350 \\
d_5 &=& \frac{{1.02342} \beta^2}{r_0^2}-\frac{{0.48610} \beta}{r_0}+{0.058569} \\
d_7 &=& \frac{{0.75158} \beta^3}{r_0^3}-\frac{{0.55570} \beta^2}{r_0^2}
+\frac{{0.13858} \beta}{r_0}-{0.01166} \\
d_9 &=& \frac{{1.15417} \beta^4}{r_0^4}-\frac{{1.18963} \beta^3}{r_0^3}+
\frac{{0.46370} \beta^2}{r_0^2}-\frac{{0.08094} \beta}{r_0}+{0.00533} \ .
\eeq

Expanding this equation in powers of $\beta$ and $\gamma$ we obtain
\beq
\Delta\phi^{(4)} \approx \frac{4\beta}{r_0}  - \gamma r_0 + \left(
\frac{15 \pi  \beta ^2}{4 r_0^2}-\frac{4 \beta ^2}{r_0^2}+\gamma  \beta -\frac{3 \
\gamma  \pi  \beta }{2}+\frac{\gamma ^2 r_0^2}{2} \right) + \dots  \ ,
\eeq
where the term between parenthesis contains the contribution of second order in the
expansion. Notice that the first two terms are those usually reported in the
literature~\cite{Ed98,Pir04b}.

\begin{figure}
\begin{center}
\includegraphics[width=9cm]{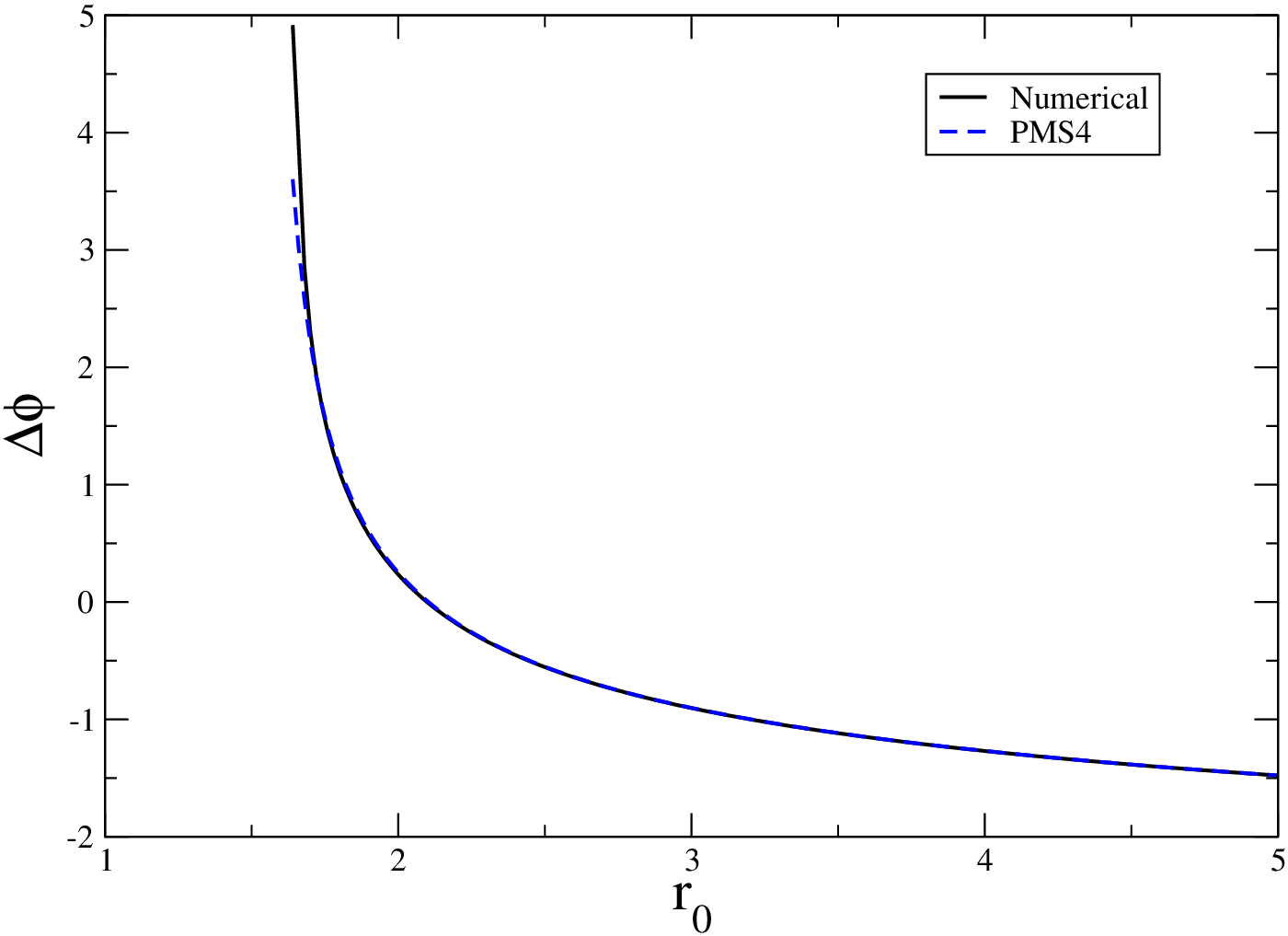}
\bigskip
\caption{Deflection angle for the Weyl metric with $\beta = \gamma = 1$ for $r_0$ close
to the photon sphere.
(color online)}
\bigskip
\label{FIG4}
\end{center}
\end{figure}

Fig.~\ref{FIG4} shows results for the deflection angle given by
the Weyl metric with $\beta = \gamma = 1$. Our fourth--order analytical formula agrees
remarkably well with the numerical calculation.

Notice that the deflection angle displays an interesting behavior depending upon the
sign of $\gamma$: as we see in Fig.~\ref{FIG4} for $\gamma>0$ the angle is negative
at large distances while it becomes positive close to the photon
sphere. Using our analytical formula we are able to provide an accurate expression
for the value of $r_0$ corresponding to absence of deflection:
\beq
r_\star \approx 2 \sqrt{\frac{\beta}{\gamma}} + \frac{\beta  \left(-\beta  (40960+9 \pi  (112+75 \pi  (-16+3 \pi ))) \gamma \
+864 \sqrt{\beta } \pi  (-16+5 \pi ) \sqrt{\gamma }+1024 (32-9 \pi)\right)}{32768} \ .
\eeq
In the case of Fig.~\ref{FIG4} this formula predicts the location of $r_\star \approx 2.08$,
which is remarkably close to the exact numerical value.

In the opposite regime, $\gamma < 0$, the deflection angle is real only on a finite region
that we obtain from
\beq
\lambda_{PMS} = 1 - \frac{8 \beta}{\pi r_0} + \frac{2 \gamma r_0}{\pi}  \ .
\eeq
In fact, when $\beta>0$ and $\gamma > - \frac{\pi^2}{64 \beta}$
then $\lambda_{PMS}>0$ only for
\beq
\frac{\sqrt{64 \beta  \gamma +\pi ^2}}{4 \gamma }-\frac{\pi }{4 \gamma } < r_0 <
-\frac{\sqrt{64 \beta  \gamma +\pi ^2}}{4 \gamma }-\frac{\pi }{4 \gamma } \ .
\eeq

\begin{figure}
\begin{center}
\includegraphics[width=9cm]{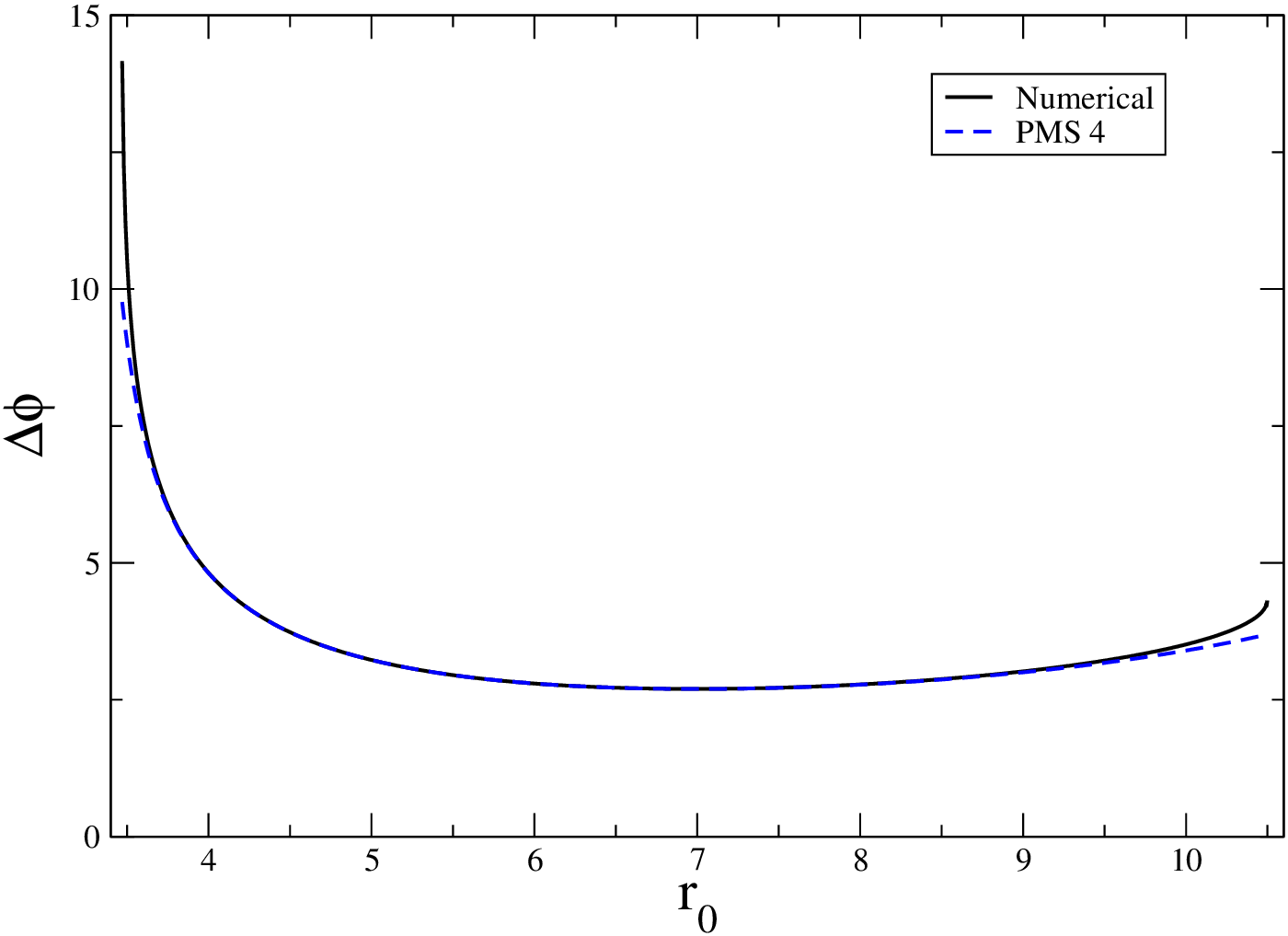}
\bigskip
\caption{Deflection angle for the Weyl metric with $\beta = 1$ and $\gamma = - \pi^2/128$.}
\bigskip
\label{FIG5}
\end{center}
\end{figure}

In Fig.~\ref{FIG5} we compare our fourth--order analytical result with the numerical
one for the Weyl metric with $\beta = 1$ and $\gamma = - \pi^2/128$. The agreement is
remarkable for all the values of $r_0$ considered.

\section{Conclusions}
\label{concl}

In this paper we have generalized a method developed earlier \cite{AA06} for obtaining
analytical expressions for the deflection angle of light travelling in an arbitrary static
and spherically symmetric metric. We have applied a new higher--order formula to
several examples confirming the accuracy suggested by a rigorous proof and estimate
of the rate of convergence \cite{AA06}.

We believe that our formalism is particularly useful for several reason: first of all,
because it stands on a firm mathematical ground; second, because it is completely
general and can be applied with little or no effort to any such metric; finally
because the analytic results obtained with our method could provide a valuable tool
for the analysis of different models. Just to mention one example, in the case of
the Weyl metric, we have been able to derive an analytical expression for the absence
of deflection for a ray of light travelling in a metric with $\beta>0$ and $\gamma>0$.

\appendix

\section{$\Omega^{(j)}_{n_1 \dots n_j}$}

Explicit expressions of the symmetric tensor $\Omega$ through order four,
which one obtains by straightforward algebra:
\beq
\Omega^{(0)} &=& \frac{\pi}{2}\\
\Omega_n^{(1)} &=& \sqrt{\pi} \frac{\Gamma(n/2+1/2)}{\Gamma(n/2)} \\
\Omega_{n_1 n_2}^{(2)} &=& \frac{2}{3} \sqrt{\pi } \left(-\frac{\Gamma \
\left(\frac{n_2+1}{2}\right)}{\Gamma \left(\frac{n_2}{2}-1\right)}-\frac{\Gamma \
\left(\frac{n_1+1}{2}\right)}{\Gamma \left(\frac{n_1}{2}-1\right)}+\frac{\Gamma \
\left(\frac{1}{2} (n_1+n_2+1)\right)}{\Gamma \left(\frac{1}{2} (n_1+n_2-2)\right)}\right) \\
\Omega_{n_1 n_2 n_3}^{(3)} &=&  \frac{4}{15} \sqrt{\pi } \left(
\frac{\Gamma \left(\frac{n_1+1}{2}\right)}{\Gamma \left(\frac{n_1}{2}-2\right)} +
\frac{\Gamma \left(\frac{n_2+1}{2}\right)}{\Gamma \left(\frac{n_2}{2}-2\right)}
+\frac{\Gamma \ \left(\frac{n_3+1}{2}\right)}{\Gamma \left(\frac{n_3}{2}-2\right)}
-\frac{\Gamma \left(\frac{1}{2} (n_1+n_2+1)\right)}{\Gamma \left(\frac{1}{2} (n_1+n_2-4)\right)}
\right.\nonumber \\
&-& \left. \frac{\Gamma \left(\frac{1}{2} \
(n_1+n_3+1)\right)}{\Gamma \left(\frac{1}{2} \
(n_1+n_3-4)\right)}-\frac{\Gamma \left(\frac{1}{2} \
(n_2+n_3+1)\right)}{\Gamma \left(\frac{1}{2} \
(n_2+n_3-4)\right)}+\frac{\Gamma \left(\frac{1}{2} \
(n_1+n_2+n_3+1)\right)}{\Gamma \left(\frac{1}{2} \
(n_1+n_2+n_3-4)\right)}\right) \\
\Omega_{n_1 n_2 n_3 n_4}^{(4)} &=&  \frac{8}{105} \sqrt{\pi } \left(
-\frac{\Gamma \left(\frac{n_1+1}{2}\right)}{\Gamma \left(\frac{n_1}{2}-3\right)}
-\frac{\Gamma\left(\frac{n_2+1}{2}\right)}{\Gamma\left(\frac{n_2}{2}-3\right)}-
\frac{\Gamma \left(\frac{n_3+1}{2}\right)}{\Gamma \left(\frac{n_3}{2}-3\right)}
-\frac{\Gamma \left(\frac{n_4+1}{2}\right)}{\Gamma \left(\frac{n_4}{2}-3\right)} \right. \nonumber \\
&-& \left. \frac{\Gamma \left(\frac{1}{2} (n_1+n_2+1)\right)}{\Gamma \left(\frac{1}{2} (n_1+n_2-6)\right)} 
          +\frac{\Gamma \left(\frac{1}{2} (n_1+n_3+1)\right)}{\Gamma \left(\frac{1}{2} (n_1+n_3-6)\right)}
          +\frac{\Gamma \left(\frac{1}{2} (n_2+n_3+1)\right)}{\Gamma \left(\frac{1}{2} (n_2+n_3-6)\right)} \right. \nonumber \\
&+& \left. \frac{\Gamma \left(\frac{1}{2} (n_1+n_4+1)\right)}{\Gamma \left(\frac{1}{2} (n_1+n_4-6)\right)} 
+ \frac{\Gamma \left(\frac{1}{2} (n_2+n_4+1)\right)}{\Gamma \left(\frac{1}{2} (n_2+n_4-6)\right)}
+\frac{\Gamma \left(\frac{1}{2} (n_3+n_4+1)\right)}{\Gamma \left(\frac{1}{2} (n_3+n_4-6)\right)}
\right.  \nonumber \\
&-& \left. \frac{\Gamma \left(\frac{1}{2} (n_1+n_2+n_3+1)\right)}{\Gamma \left(\frac{1}{2} (n_1+n_2+n_3-6)\right)}  
          -\frac{\Gamma \left(\frac{1}{2} (n_1+n_2+n_4+1)\right)}{\Gamma \left(\frac{1}{2} (n_1+n_2+n_4-6)\right)} \right. \nonumber \\
&-& \left. \frac{\Gamma \left(\frac{1}{2} (n_1+n_3+n_4+1)\right)}{\Gamma \left(\frac{1}{2} (n_1+n_3+n_4-6)\right)}
-\frac{\Gamma \left(\frac{1}{2} (n_2+n_3+n_4+1)\right)}{\Gamma \left(\frac{1}{2} (n_2+n_3+n_4-6)\right)}
+\frac{\Gamma \left(\frac{1}{2} (n_1+n_2+n_3+n_4+1)\right)}{\Gamma \left(\frac{1}{2} (n_1+n_2+n_3+n_4-6)\right)}\right)   \ .
\eeq

\section{$\Sigma_k$}

Explicit expressions for the coefficients $\Sigma_k$ defined
in Eq.~(\ref{sigma}) for the Taylor expansion of an arbitrary potential
through order three:
\beq
\Sigma_0 &=& \frac{\pi }{2} \\
\Sigma_1 &=& v_1+\frac{1}{2} \pi  v_2+2 v_3 + \dots \\
\Sigma_2 &=& \frac{2 v_1^2}{3}+2 v_2 v_1+\pi  v_3 v_1-\frac{2}{3} v_3 \
v_1+\frac{1}{2} \pi  v_2^2 \nonumber \\
&+&\frac{5}{4} \pi  v_3^2-\frac{4 v_3^2}{3}+4 \ v_2 v_3 +\dots \\
\Sigma_3 &=& \frac{7 v_1^3}{15}+2 v_2 v_1^2+\frac{12}{5} v_3 v_1^2+3 v_2^2 \
v_1-\frac{3}{2} \pi  v_3^2 v_1 \nonumber \\
&+&\frac{47}{5} v_3^2 v_1+3 \pi  v_2 v_3 \
v_1-2 v_2 v_3 v_1+\frac{1}{2} \pi  v_2^3-\frac{3}{2} \pi  \
v_3^3 \nonumber \\
&+&\frac{122 v_3^3}{15}+\frac{15}{4} \pi  v_2 v_3^2-4 v_2 v_3^2+6  v_2^2 v_3 +\dots  \\
\Sigma_4 &=&  \frac{12 v_1^4}{35}+\frac{28}{15} v_2 v_1^3+\frac{76}{35} v_3 v_1^3+4 \
v_2^2 v_1^2+3 \pi  v_3^2 v_1^2 \nonumber \\
&-& \frac{124}{35} v_3^2 v_1^2+\frac{48}{5} \ v_2 v_3 v_1^2+4 v_2^3 v_1+9 \pi  v_3^3 v_1 \nonumber \\
&-& \frac{708}{35} v_3^3 \ v_1-6 \pi  v_2 v_3^2 v_1+\frac{188}{5} v_2 v_3^2 v_1\nonumber \\
&+& 6 \pi  v_2^2 \ v_3 v_1-4 v_2^2 v_3 v_1+\frac{1}{2} \pi  v_2^4+\frac{99}{16} \pi  \
v_3^4-\frac{104 v_3^4}{7} \nonumber \\
&-&6 \pi  v_2 v_3^3+\frac{488}{15} v_2 \
v_3^3+\frac{15}{2} \pi  v_2^2 v_3^2-8 v_2^2 v_3^2+8 v_2^3 v_3 + \dots
\eeq

\begin{acknowledgments}
P.A. acknowledges support of Conacyt grant no. C01-40633/A-1.
\end{acknowledgments}


\begin{thebibliography}{}
\bibitem{AA06} P.Amore and S.Arceo, Phys.Rev.{\bf D} 73, 083004 (2006)
\bibitem{FKN00} S. Frittelli, T.P. Kling and T.Newman, Phys.Rev.{\bf D} 61, 064021 (2000)
\bibitem{VE00} K.S.Virbhadra and G.F.R.Ellis,Phys. Rev. {\bf D} 62, 084003 (2000)
\bibitem{VE02} K.S.Virbhadra and G.F.R.Ellis,Phys. Rev. {\bf D} 65, 103004 (2002)
\bibitem{VNC98} K.S. Virbhadra, D. Narasimha and S.M.Chitre, Astron. Astrophys.337,1-8 (1998)
\bibitem{Eir02} E.F.Eiroa,G.E.Romero and D.F.Torres, Phys. Rev. {\bf D} 66, 024010 (2002)
\bibitem{Bha03} A. Bhadra, Phys. Rev. {\bf D} 67, 103009 (2003)
\bibitem{Boz03} V. Bozza, Phys. Rev. {\bf D} 67, 103006 (2003);
                V. Bozza, F. De Luca, G. Scarpetta, M. Sereno, Phys.Rev. {\bf D}72, 08300 (2005);
                V. Bozza, F. De Luca and G. Scarpetta, arXiv:gr-qc/0604093.
\bibitem{Whi05} R. Whisker, Phys. Rev. {\bf D} 71, 064004 (2005)
\bibitem{Eir05} E.F. Eiroa, Phys. Rev. {\bf D} 71, 083010 (2005)
\bibitem{Eir06} E.F.Eiroa, Phys. Rev. {\bf D} 72, 043002 (2006)
\bibitem{SB06} K. Sarkar and A. Bhadra, ArXiv:[gr-qc/0602087] (2006)
\bibitem{Perl04} V. Perlick, Phys.Rev. {\bf D} 69, 064017 (2004)
\bibitem{Boz02} V.Bozza, Phys. Rev. {\bf D} 66, 103001 (2002)
\bibitem{Mutkaa} P.T. Mutka and P. M\"ah\"onen, The Astrophysical Journal {\bf 581}: 1328-1336 (2002)
\bibitem{Mutkab} P.T. Mutka and P. M\"ah\"onen, The Astrophysical Journal {\bf 576}: 107-112 (2002)
\bibitem{Belo02} A.M. Beloborodov, The Astrophysical Journal {\bf 566}: L85-L88 (2002)
\bibitem{Keet05} C.R. Keeton and A.O. Petters, Phys. Rev. {\bf D} 72, 104006 (2005)
\bibitem{Weinberg}       S. Weinberg, Gravitation and cosmology, J.Wiley and Sons, 1972
\bibitem{Am05a} P.Amore and R.A.Sa\'enz, Europhysics letters {\bf 70} 425-431 (2005)
\bibitem{Am05b} P.Amore, A.Aranda, F.Fernandez and R.A.Sa\'enz, Phys. Rev.{\bf E} 71 (2005)
\bibitem{Ste81} P.M. Stevenson, Phys. Rev. D {\bf 23}, 2916 (1981)
\bibitem{Darw59} C. Darwin, Proc.R. Soc. London {\bf A} 249, 180 (1959); C. Darwin, Proc.R. Soc. London {\bf A} 263, 39 (1961);
\bibitem{MK89} P.D. Mannheim and D. Kazanas, Astrophys. J. {\bf 342}, 635-638 (1989)
\bibitem{Ed98} A. Edery and M.B.Paranjape, Phys. Rev.{\bf D} 58, 1-8 (1998)
\bibitem{Pir04a} S. Pireaux, Class. Quantum Grav. {\bf 21}, 1897-1913 (2004)
\bibitem{Pir04b} S. Pireaux, Class. Quantum Grav. {\bf 21}, 4317-4333 (2004)



\end{thebibliography}
\end{document}